\documentstyle[amssymb,prb,twocolumn,aps,epsfig]{revtex} 
  
\begin{document}  
 
\wideabs{       
\title{Phase diagram of the Hubbard chain with two atoms per cell}  
\author{M.E. Torio$^{a}$, A.A. Aligia$^{b}$ and H.A. Ceccatto$^{a}$}  
\address{$^{a}$Instituto de F\'{\i}sica Rosario, Consejo Nacional de  
Investigaciones Cient\'{\i}ficas y T\'ecnicas \\ 
and Universidad Nacional  
de Rosario, Boulevard 27 de Febrero 210 bis, (2000) Rosario, Argentina \\  
$^{b}$Centro At\'{o}mico Bariloche and Instituto Balseiro, \\ 
Comisi\'on Nacional de Energ\'{\i}a At\'{o}mica,  
8400 Bariloche, Argentina}  
\maketitle  
  
\begin{abstract}  
We obtain the quantum phase diagram of the Hubbard chain with alternating  
on-site energy at half filling. The model is relevant for the ferroelectric  
perovskites and organic mixed-stack donor-acceptor crystals. For any values  
of the parameters, the band insulator is separated from the Mott insulator  
by a dimer phase. The boundaries are determined accurately by crossing of  
excited levels with particular discrete symmetries. We show that these  
crossings coincide with jumps of charge and spin Berry phases with a clear  
geometrical meaning.   
\end{abstract}  
  

}  
 
\newpage  
  
The transition between a band insulator and a Mott insulator in a  
one-dimensional (1D) model for ferroelectric perovskites \cite{ega} has been  
a subject of great interest in recent years  
\cite{ega,res1,ort,res2,fab,gid,taka,qin,wil}. The model describing this  
transition, originally  
proposed \cite{nag} for the neutral-ionic transition in mixed-stack  
donor-acceptor organic crystals, \cite{tor,avi} is 
\begin{eqnarray}  
H &=& -t\sum_{i\sigma }(c_{i+1\sigma }^{\dagger }c_{i\sigma }+{\rm H.c.}) 
+\Delta \sum_{i\sigma }(-1)^{i}n_{i\sigma } \nonumber \\ 
& & +U\sum_{i}n_{i\uparrow }n_{i\downarrow }.  
\label{h}  
\end{eqnarray}  
At a fixed value of $\Delta $, exact-diagonalization studies on rings of up  
to 12 sites \cite{res1,gid} and Hartree-Fock calculations \cite{ort} found  
evidence of a transition with increasing $U$ from a band insulating (BI)  
ionic phase to a Mott insulating (MI) quasi neutral phase, which could be  
expected on general grounds. Furthermore, the transition point was  
characterized \cite{res2} as a metallic point, with divergent  
delocalization. On the other hand, a field theoretical approach, \cite{fab}  
valid in the weak coupling limit $(\Delta ,U) \ll t$, concluded that a  
spontaneously dimerized insulating (SDI) phase (also called bond-ordered  
wave) intervenes between the BI and MI phases. However, due to the  
limitations of this technique the precise extension of this phase remained  
unknown. Very recently, density matrix renormalization group (DMRG)  
investigations \cite{taka,qin} and a quantum Monte Carlo (QMC) approach \cite  
{wil} found contradictory evidence: the two DMRG calculations reached  
opposite conclusions regarding the existence of the SDI phase, while in the  
QMC results the SDI-MI phase transition was not observed. Thus, the  
existence of all these conflicting results calls for a further investigation  
of this model.  
  
In this Letter we clarify this controversy and accurately determine the whole  
ground-state phase diagram of Hamiltonian Eq. (\ref{h}).  
This is accomplished by the combined use of the method of topological  
transitions (MTT) (jumps in charge and spin Berry phases) \cite{gag,ali,top} 
and the method of crossing excitation levels(MCEL)
based on the conformal field theory with renormalization group  
analysis \cite{nom,nak}. These methods, briefly explained below, are somehow  
complementary in the sense that while the geometrical content of the MTT is  
more clearly displayed in the strong coupling limit, the MCEL is based on a  
weak-coupling approach. A nice feature here is that they turn out to be  
equivalent for this problem, so that the results obtained are expected to be  
valid for all parameter values. State of the art diagonalization of rings  
with up to 16 sites are performed to determine the phase boundaries with  
errors estimated in a few percent of $t$.  
  
The Berry phases are calculated numerically from the ground state $|g(\Phi  
_{\uparrow },\ \Phi _{\downarrow })\rangle $\ of $\tilde{H}(\Phi _{\uparrow  
},\ \Phi _{\downarrow })$ in rings of even number of sites $L$ threaded by  
fluxes $\Phi _{\sigma }$~for spin $\sigma $. The Hamiltonian $\widetilde{H}$  
differs from $H$ in that the hopping term has the form $-t\sum_{i\sigma }(%
\widetilde{c}_{i+1\sigma }^{\dagger }\widetilde{c}_{i\sigma }e^{i\phi  
_{\sigma }/L}+{\rm H.c.}).$ One can map $\widetilde{H}$ with periodic  
boundary conditions (BC) into $H$ with twisted BC ($c_{i+L\sigma }^{\dagger  
}=e^{i\phi _{\sigma }}c_{i\sigma }$) using the canonical transformation  
$c_{j\sigma }=e^{ij\phi _{\sigma }/L}\widetilde{c}_{j\sigma }$.   
The charge (spin) Berry phase $\gamma _{c\text{ }}$  
($\gamma _{s}$) is the phase captured by the ground state when it is followed  
adiabatically in the cycle $0\leq \Phi \leq 2\pi $, keeping $\Phi _{\uparrow  
}=\Phi _{\downarrow }=\Phi $ ($\Phi _{\uparrow }=-\Phi _{\downarrow }=\Phi $%
). Discretizing the interval $0\leq \Phi \leq 2\pi $ into $N+1$ points $\Phi  
_{r}=2\pi r/N$ ($r=0,N$), the Berry phases are calculated using: \cite{gag}   
\begin{eqnarray}  
\gamma _{c,s} &=&-\lim_{N \rightarrow \infty } \text{Im} \{ \ln [ 
\prod^{N-2}_{r=0} \langle g(\Phi _{r},\pm \Phi _{r})|g(\Phi  
_{r+1},\pm \Phi _{r+1}) \rangle  \nonumber \\  
& & \times \langle g(\Phi _{N-1},\pm \Phi _{N-1})|g(2\pi,  
\pm 2\pi) \rangle ]  \},  \label{gamma}  
\end{eqnarray}  
\noindent where $|g(2\pi ,\pm \ 2\pi )\rangle = \exp [i{\frac{2\pi}{L}} 
\sum_{j} j( n_{j\uparrow } \pm n_{j\downarrow })] |g(0,0) \rangle$.   
  
An important property of $\gamma _{c}$ is that if the system is  
modified by some perturbation, the change in the polarization $P_{\uparrow  
}+P_{\downarrow }$ is proportional to the corresponding change in $\gamma  
_{c}.$\cite{ort} Here $P_{\sigma }$ is the contribution of electrons with  
spin $\sigma $ to the polarization of the system. Similarly, changes in $%
\gamma _{s}$ are related to changes in the difference $P_{\uparrow  
}-P_{\downarrow }$ between the electric polarizabilities for spins up and  
down: \cite{ali}  
$\Delta P_{\uparrow }\pm \Delta P_{\downarrow }=e\Delta \gamma _{c,s}/2\pi   
[{\rm mod}(e)]$.   
A more crucial property is that in systems with inversion symmetry $\gamma  
_{c}$ and $\gamma _{s}$ can only be either $0$ or $\pi $, which has  
led to the idea that $\overline{\gamma }=(\gamma _{c},\gamma _{s})$ can be  
used as a topological vector to characterize different phases. \cite{gag}  
Such a possibility is clear in the strong-coupling limit $t\rightarrow 0$,  
where (usually) all particles are localized: one can choose a gauge in which  
all scalar products in Eq. (\ref{gamma}) except the last one are equal to $1$%
, so that $\gamma _{c}$ is determined by the sum  
$i{\frac{2\pi }{L}}\sum_{j=0}^{L-1}{j(}n_{j\uparrow}+n_{j\downarrow })$. 
For example, if there is one particle per site ($U\rightarrow  
\infty $), it gives $i\pi (L-1)\equiv i\pi $ $[{\rm mod}(2\pi i)]$ for $L$  
even, and then $\gamma _{c}=\pi .$ Similarly, for a N\'{e}el state it is  
easy to see that $\gamma _{s}=\pi $, and for a charge density wave (CDW)  
with maximum order parameter ($\Delta \rightarrow \infty $) $\overline{%
\gamma }=(0,0)$. These values are consistent with the changes in $%
P_{\uparrow }\pm P_{\downarrow }$ originated by the  
charge transport of all electrons with a given spin to nearest-neighbor  
sites, required to change the extreme BI state with $\overline{\gamma }%
=(0,0) $ to the N\'{e}el state with $\overline{\gamma }=(\pi ,\pi ).$ For  
the extreme MI state ($U\rightarrow \infty $), which is a spin-density wave  
(SDW), we also have $\overline{\gamma }=(\pi ,\pi )$. \cite{gag,top} By  
continuity, one might expect that these values of $\overline{\gamma }$  
characterize also the BI and MI phases in weak coupling. As explained below,  
this is confirmed by an analysis based on the MCEL. 
This change in topological parameters (which is sharp even in finite systems) indicates  
non-trivial changes in $P_{\uparrow }\pm P_{\downarrow }$ characteristic of  
a phase transition.  
 
We find another phase  
with $\overline{\gamma }=(\pi ,0)$. From field theory \cite{fab} we  
know that this corresponds to the SDI phase with order parameter  
$D=\sum_{j\sigma }(-1)^{j}(c_{j+1\sigma }^{\dagger }c_{j\sigma }+{\rm H.c.}).$  
If we consider the more general Hamiltonian  
\begin{eqnarray}  
H^{\prime }&=&H-(t_{AB}-t)\sum_{i\sigma }(c_{i+1\sigma }^{\dagger }c_{i\sigma  
}+{\rm H.c.})(n_{i\overline{\sigma }-}n_{i+1\overline{\sigma }})^{2} \nonumber\\ 
&&+\;V\sum_{i\sigma \sigma ^{\prime }}n_{i\sigma }\ n_{i+1\sigma^{\prime }},   
\label{hp}  
\end{eqnarray}  
we confirm that the SDI phase of $H^{\prime }$, well established in previous  
studies, \cite{top,nak,jap,bos,pola} is smoothly connected with that of $H$  
for $t_{AB}\rightarrow \infty $. Furthermore, the model of Eq. (\ref{hp})  
with $\Delta =0$ and $(V,t_{AB}-t)>0$ contains essentially the same phases  
as $H$, and allows a more detailed study of the relation between the MTT and  
MCEL. For $V=0$, while DMRG results in chains of 40 sites are unable  
to detect the opening of an exponentially small gap,\cite{top}  
the MTT predictions with $L$ up to 12 \cite{top} practically coincide  
with those of field theory for $t_{AB}\sim 1$ \cite{jap,bos} and with  
exact results for $t_{AB}\rightarrow 0$.\cite{afq} 
  
The MCEL is based on the fact that in a conformal field theory (which  
ultimately describes the low-energy physics of 1D systems in the charge and  
spin sectors if they are gapless) the exponent $\nu $ of the long-distance  
power-law decay of correlation functions $\left\langle  
O(x)O(x+d)\right\rangle \sim d^{-\nu }$ is given in terms of excitation  
energies related to the operator $O(x)$ in the finite ring. A crossing of  
appropriately chosen excitation energies for different operators indicates a  
change in the character of the dominant correlations at large distance (a  
phase transition). \cite{nom,nak} The relevant excitation energies for $%
H^{\prime }\ $with $\Delta =0$ have been studied by Nakamura. \cite{nak} In  
particular, in the weak coupling limit it is known that there is a Gaussian  
transition from the CDW to the SDI in $H^{\prime }\ (\Delta =0)$, with the  
charge gap vanishing only at the transition point.\cite{nak,jap,bos} This  
transition is determined by the crossing of the lowest states with opposite  
parity under inversion and the same total momentum $K=\pi /a$ ($a$ is the  
nearest-neighbor distance), calculated with periodic (antiperiodic) BC if $%
L=4n$ ($4n+2$). \cite{nak} We show that this crossing coincides with the  
jump in $\gamma _{c}.$ In Fig. 1 we represent $E(\Phi )=\langle g(\Phi ,\Phi  
)|\widetilde{H}^{\prime }|g(\Phi ,\Phi )\rangle $ for the simplest case with   
$L$ multiple of four. Minimizing $E(\Phi )$with respect to $\Phi $ and $%
\widetilde{K}$ leads to $\Phi =\pi ,\ \widetilde{K}=\pi /a$, and the Berry  
phases are obtained following adiabatically this state. Using  
$c_{j\sigma }=e^{ij\phi _{\sigma }/L}\widetilde{c}_{j\sigma }$,  
one sees that while the total wave vector  
$\widetilde{K}$ of $\widetilde{H}^{\prime }$ remains constant  
as $\Phi $ is changed, in general $K=\widetilde{K}+(N_{\uparrow } 
\Phi _{\uparrow }+N_{\downarrow }\Phi  
_{\downarrow })/(L\ a)$, where $N_{\sigma }$ is the number of particles with  
spin $\sigma $. The conditions leading to the minimum energy ($\Phi =\pi ,\   
\widetilde{K}=\pi /a$) correspond to antiperiodic BC and $K=0$ in $H^{\prime  
}$.  

\vspace{1.13cm} 

\begin{figure}[tbp]
\epsfxsize=6cm
\epsfysize=10.cm
\centerline{\epsfbox{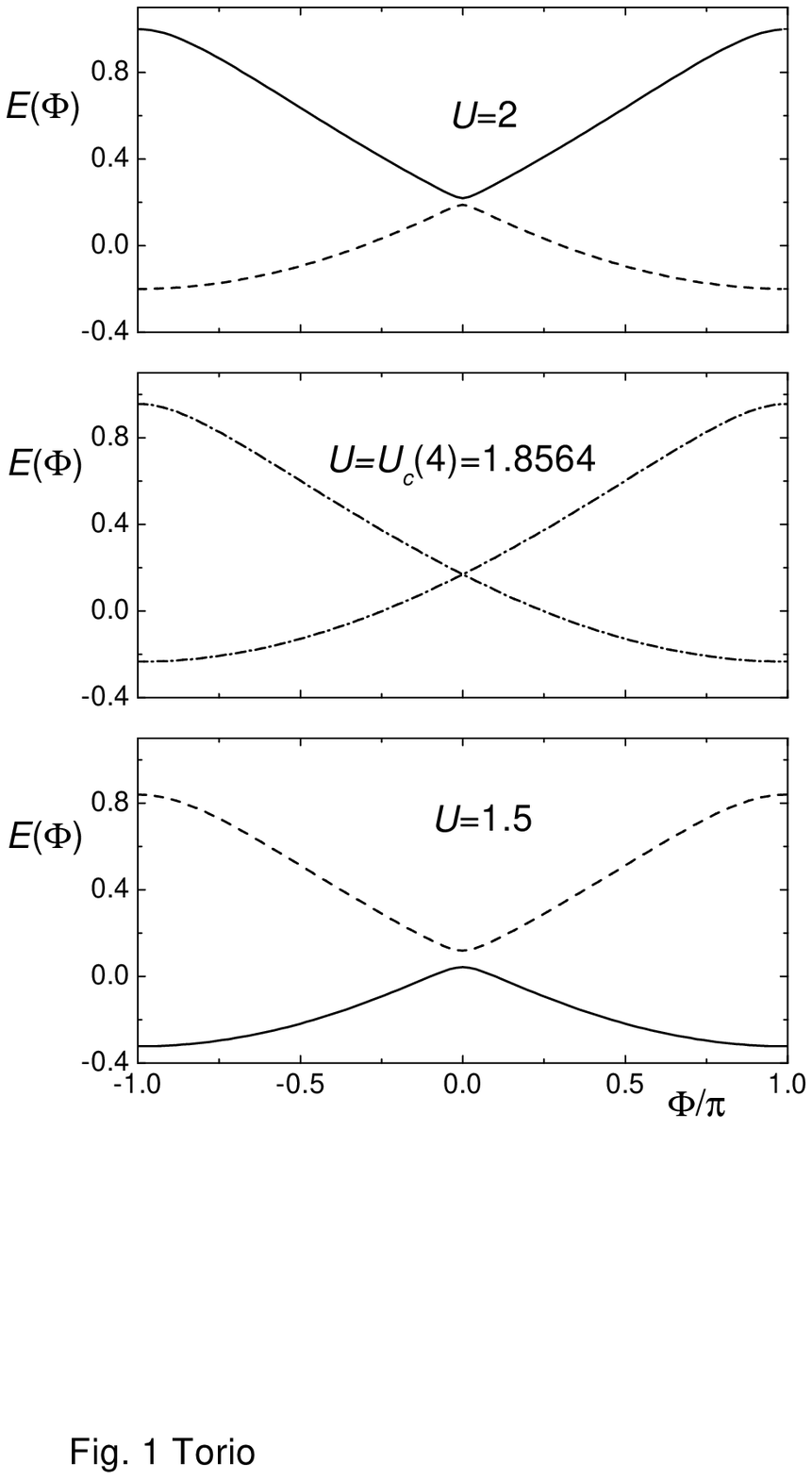}}
\vspace{-1.2cm}

\narrowtext
\caption{Energy per site as a function of flux $\Phi _{\uparrow }=\Phi   
_{\downarrow }=\Phi$ for the two lowest lying eigenstates within the   
subspace of $\widetilde{K}=\pi /a,\ S=0$. Parameters are $L=4$, $\Delta=0$,   
$t_{AB}=t=1$, $V=2$   
and three different values of $U$ as indicated. Full (dashed) line   
correspond to states with $\gamma _{c}=0\ (\gamma _{c}=\pi )$. For   
$U=U_{c}(L) $ $\gamma _{c}$ is undefined due to the degeneracy at $\Phi =0.$  }.
\end{figure}

\vspace{-.2cm}   
It is easy to see that the inversion $I\,c_{j\sigma }^{\dagger }\,I^{\dagger  
}=c_{-j\sigma }^{\dagger }\equiv e^{-i\Phi _{\sigma }}c_{L-j\sigma  
}^{\dagger }$ is a symmetry of $H^{\prime }(\Phi _{\uparrow },\Phi  
_{\downarrow })$ only if both $\Phi _{\sigma }$ are either $0$ or $\pi $  
(corresponding to periodic or antiperiodic BC). At $U=U_{c}(L)$ and $\Phi  
=0$ there is a crossing of the lowest levels with $\ \widetilde{K}=\pi /a$  
and total spin $S=0.$ This crossing is possible because the corresponding  
wave functions have opposite parity, and therefore they do not mix at $\Phi  
=0.$ For $\Phi \rightarrow 0$ one can use perturbation theory in $\Phi $, and%
$\ $for $U$ near $U_{c}(L)$ the state $|g(\Phi )\rangle $ is determined by a   
$2\times 2$ matrix involving the above mentioned two states for $\Phi =0$  
with off-diagonal matrix elements linear in $\Phi $. From the trivial  
solution to this problem one realizes that the product in Eq. (\ref{gamma})  
for $U\rightarrow U_{c}(L)-0^{+}$ differs from that for $U\rightarrow  
U_{c}(L)+0^{+}$ in sign. Hence, $\gamma _{c}$ jumps at the same place where  
the transitions occurs according to the MCEL. While for the $t-U-V$  
model ($H^{\prime }$ with $\Delta =0=t_{AB}-t$) in weak coupling the  
transition is second order and the charge gap vanishes at the transition,  
for  $(U,V) \gg t$ the transition is first order, from a fully gapped CDW to a  
charge gapped SDW.\cite{nak,pola,hir} As a consequence, the MCEL looses its support from the  
conformal invariant (massless) theory. However, in this limit the  
geometrical meaning of the jump in $\gamma _{c}$ is very clear,  
as explained earlier, and justifies the method.  
  
In the MCEL, the Kosterlitz-Thouless transition, which corresponds to the  
opening of a spin gap, is detected through the crossing of a singlet even  
under inversion with an odd triplet, with $K=0$ and periodic BC (for $L=4n$).  
\cite{nak} At $\Phi =0$ or $\Phi =\pi $, $H^{\prime }(\Phi ,-\Phi )$ has  
SU(2) and inversion symmetries, which are lost for other values of $\Phi $.  
Therefore, a similar analysis as above shows that $\gamma _{s}$ jumps at  
this point. \cite{ali} If $L=4n+2$, periodic and antiperiodic BC are  
interchanged. When $\Delta \neq 0$ the symmetry under translations in one  
lattice spacing $a$ is lost, $K=\pi /a$ becomes equivalent to $K=0$, and the  
CDW order parameter is different from zero also in the SDI and MI phases.  
The field theory results for $H $ show that for $(\Delta ,U)\ll t $ the spin  
transition retains the same features.\cite{fab} The charge  
transition, which for $\Delta =0$ is described by the sine Gordon model, for   
$\Delta \neq 0$ is determined by the double sine Gordon  
model, and the universality class changes from Gaussian to Ising. However,  
the transition remains second order and the charge gap vanishes at the  
transition.\cite{fab} Then, at this point and sufficiently low energies the  
charge sector is described by a conformal invariant theory, justifying  
the MCEL.   
  
In spite of the breaking of traslational symmetry, fortunately the relevant  
crossings for $H$ can still be identified looking for the ground state  
energy in subspaces with $K=0$, total spin projection $S_{z}=0$, and a  
definite parity under inversion and time reversal. The latter allows us to  
separate states with even and odd $S$. If the more general model $H^{\prime }$ 
with $\Delta \neq 0$ is  
considered, there are some regions of parameters in which the charge  
transition corresponds to a crossing of first excited states within the  
above mentioned subspaces, but we restrict ourselves here to the phase  
diagram of $H.$ For this model, the connection between the jump in $\gamma  
_{c}$ and a symmetry switch of the ground state for appropriate BC has been  
noted earlier, \cite{res1,ort,gid} but the relation with the MCEL has not  
been discussed. Moreover, neither results for $\gamma _{s}$ nor numerical  
investigation of the SDI-MI transition has been reported so far. The  
calculation of $\gamma _{s}$ in $H$ presents technical difficulties due to  
additional crossing of levels (not related with phase transitions) which  
take place for $\widetilde{K}=0.$ We have verified numerically that the  
jumps in $\gamma _{c}$ and $\gamma _{s}$ correspond to the above  
mentioned level crossings.  
  
For given $\Delta ,$ we have calculated the critical on-site repulsion $%
U_{c} $ ($U_{s}$) at which the charge (spin) transition takes place. In 
addition, for  
small $U$ and $\Delta $ we have fixed $U$ and determined the  
critical values $\Delta _{c}$ and $\Delta _{s}$. This was done by fitting a  
quadratic polynomial in $1/L^{2}$ to the results for $L=10$,$\ 12$, 14 and  
16, followed by an extrapolation to $L=\infty $. This fit works very well for   
$\Delta \geq 0.25$ (we set $t=1$ as the unit of energy unless otherwise  
stated), and improves with increasing $\Delta $. The  
difference between $U_{s}(L)$ and $U_{s}(L+2)$ rapidly decreases with $L$ if   
$\Delta $ is not too small. Instead, for small values of $\Delta $ the  
finite size effects increase and, as a consequence, the error in the  
extrapolation becomes larger. To estimate this error we have repeated  
the fits using $L=8,\ 10,\ 12$ and $14$; for $\Delta =0.05$ this gives a new 
estimation of $U_{c}$ ($U_{s}$) that differs from the previous one in $0.12$ 
($0.07$). For  
$\Delta <0.05$ the relative error in $U_{c}$ and $U_{s}$ becomes very large,  
and we do not present results since they loose quantitative validity (except  
at $\Delta =0$, where $U_{c}=U_{s}=0$ for all $L$). Instead, for $\Delta  
\geq 0.25$ the estimated error in $U_{c},U_{s}$ is less  
than 0.06, and less than 0.03 for $\Delta \geq 0.5$.  
  
The resulting phase diagram is presented in Fig. 2. In qualitative agreement  
with field theory results, \cite{fab} and for all values of $\Delta $, the  
transition from the BI phase to the MI phase with increasing $U$ takes place  
in two steps: first, a charge transition to the SDI phase at $U=U_{c}$  
occurs, and then, for $U=U_{s}>U_{c}$, the spin gap closes. The behavior in  
the strong coupling limit is quite different from that of the $t-U-V$ model,  
for which a first order CDW-SDW transition occurs and is easily  
understood in terms of perturbation theory (PT) \cite{nak,pola,hir}. Instead, $H$  
remains non-trivial for $t\rightarrow 0$ as long as $U-2\Delta \sim t$,  
since charge fluctuations are still possible. As a consequence of this  
delocalization of charges, $\gamma _{c}$ inside the SDI phase cannot be  
calculated analytically just adding the position of the charges, as we  
explained before for $\Delta \rightarrow \infty $ or $U\rightarrow \infty $.  
For $t=0$ the SDI phase is absent, and PT in $t$ diverges  
at $U=2\Delta $ where the BI-MI transition takes place. For $t\ll |U-2\Delta  
|\ll U$ PT is valid, and can be used to calculate the  
energy of the BI and MI phases for negative and positive $U-2\Delta $  
respectively. The MI phase in this limit is described by a Heisenberg model  
with exchange $J=2t^{2}/(U-2\Delta )$. The energies up to second order in $t$  
are given by  
$E_{BI}=U-2\Delta + 4t^{2}/(U-2\Delta)$ and  
$E_{MI}= - J\ln 2$.    
While the SDI phase cannot be described by PT in $t$, its  
boundaries are very accurately determined by our method for small $t$. The  
jumps in Berry phases have very little size dependence and show that 
$U_{c} \simeq 2\Delta +1.33t$ and $U_{s} \simeq 2\Delta +1.91t$ for  
$t \ll (\Delta ,U)$. The fact that the SDI phase exists for positive values  
of $U-2\Delta $ was to be expected from the asymmetry of $E_{BI}$ and  
$E_{MI}$ under a change of sign of $U-2\Delta $.  
  
The results for $(\Delta ,U)\gg t$ can be extended qualitatively to $\Delta  
\sim t.$ The SDI has a nearly constant width $\sim 0.6t$, and both $U_{s}$  
and $U_{c}$ increase slightly with decreasing $\Delta $. For $\Delta <1$ the  
critical values $U_{c}$ and $U_{s}$ decrease abruptly, until they reach $%
U_{c}=U_{s}=0$ at $\Delta =0$. However, in the region $0<\Delta <0.25$ ($%
0<U\lesssim 2$) the relative errors in $U_{c}$, $U_{s}$ become larger with  
decreasing $\Delta $; in particular, for $\Delta \lesssim 0.1$ our results  
are not quantitatively reliable. For $(\Delta ,U)\ll t$, the spin  
transition can be estimated integrating out the charge degrees of freedom,  
assuming that they are described by a free massive boson. This leads to a  
renormalization of the effective interaction $g_{1\perp }$ responsible for  
the opening of a spin gap in the sine Gordon model which describes the spin  
sector at low energies.\cite{fab} From the vanishing of the renormalized  
$g_{1\perp }$ one obtains the approximate field theory result    
$\Delta _{s}^{ft}\sim E_{g}\sqrt{U/(8\pi t)}$,   
where $E_{g}$ is the gap for $\Delta =0$ and is known from the Bethe  
ansatz solution. For $U\ll t$, $E_{g}\cong (8/\pi )\sqrt{tU}\exp (-2\pi t/U)$,  
and the exponential dependence dominates the behavior of  
$\Delta _{s}^{ft}$. Due to the numerical uncertainties for  
$\Delta \lesssim 0.1$, we cannot  
establish where this exponential dependence deviates from the actual SDI-MI  
boundary. For small $\Delta $ and any value of $U$, an accurate field theory  
result for $\Delta _{s}$ might be obtained using a bosonization approach  
which starts from the exact solution for $\Delta =0$.\cite{cabra}   

\vspace{-.4cm}

\begin{figure}[tbp]
\epsfxsize=11cm
\epsfysize=13.cm
\centerline{\epsfbox{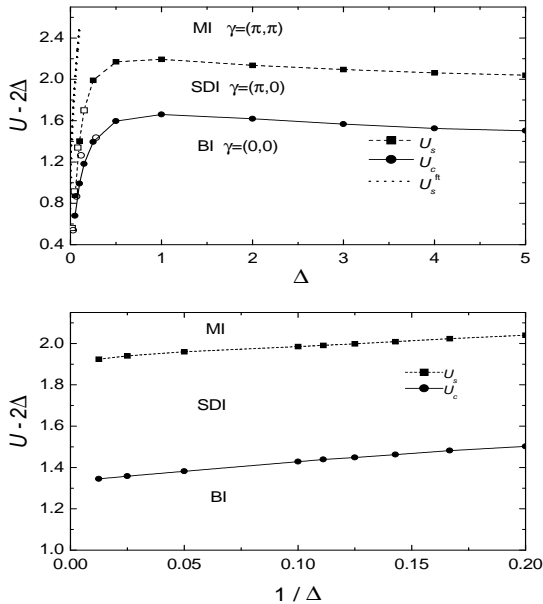}}
\vspace{-2.95cm}
\narrowtext
\caption{Ground-state phase diagram of $H$ at half filling. The dashed line
corresponds   to the field theory result $\Delta _{s}^{ft}$ (see main text).
The open (full) symbols   were obtained keeping $U$ ($\Delta$) constant. }.
\end{figure}

\vspace{-.2cm}  
The SDI order parameter $D$ couples directly with optical phonons with wave  
vector $K=0$ and, therefore, the latter should increase the extension of  
this phase. In principle, one can include these phonons in the numerical  
calculations using the adiabatic approximation. However, due to the breaking  
of inversion symmetry our method cannot be used to find the phase boundaries  
in this case. QMC calculations suggest that in the adiabatic  
approximation the whole MI phase disappears and the SDI takes its place.  
\cite{wil} This is not necessarily the case if the dynamics of  
the phonons is included.\cite{fab} We must emphasize that in the MI phase  
both dimer-dimer and spin-spin correlation functions have  
the same leading  power-law decay 
at large distances. The dominance of spin-spin correlations due to  
logarithmic corrections characterizes the MI phase.\cite{top,jap,bos} This  
renders it very difficult to determine the SDI-MI boundary   
by direct numerical evaluation of correlation functions.\cite{top}  
  
In summary, we have determined the quantum phase diagram of the Hubbard  
chain with alternating on-site energies at half filling using topological  
transitions. The method is justified from geometrical considerations in the  
strong coupling limit ($t\rightarrow 0$) and by field theory arguments in  
the weak-coupling $(U,\Delta )\ll t$ region. We confirmed the existence of a  
spontaneously dimerized phase and determined its boundaries for the first  
time.  
  
One of us (AAA) thanks D.C. Cabra for useful discussions and  
acknowledges computer time at the Max-Planck Institute f\"{u}r Physik  
Komplexer Systeme. The authors are partially supported by CONICET. This work  
was sponsored by PICT 03-00121-02153 and PICT 03-06343 from ANPCyT and PIP  
4952/96 from CONICET.

\end{document}